\documentclass[12pt, draftclsnofoot,onecolumn]{IEEEtran}
\usepackage{xcolor}
\usepackage{cite, graphicx}
\usepackage{hyperref}
\usepackage[cmex10]{amsmath}
\usepackage{amssymb}
\usepackage{siunitx}
\ifCLASSOPTIONcompsoc
\usepackage[caption=false,font=normalsize,labelfont=sf,textfont=sf]{subfig}
\else
\usepackage[caption=false,font=footnotesize]{subfig}
\fi
\usepackage{tabularx}
\usepackage{paralist}
\makeatletter
\let\savespace\@minipagetrue
\makeatother

\begin{document}
\title{Software-Defined Hyper-Cellular Architecture for Green and Elastic Wireless Access}%

\author{Sheng~Zhou, Tao Zhao, Zhisheng~Niu, and Shidong Zhou
\thanks{Sheng~Zhou, Zhisheng~Niu and Shidong Zhou are with Tsinghua National Laboratory for Information Science and Technology, Dept. of
Electronic Engineering, Tsinghua University, Beijing 100084, China.

Tao Zhao was with Tsinghua National Laboratory for Information Science and Technology, Dept. of
Electronic Engineering, Tsinghua University, Beijing, China. He is now with Department of Electrical \& Computer Engineering, Texas A\&M University, College Station, Texas, United States.

This work is sponsored in part by the National Basic Research Program of China (973 Program: No. 2012CB316000), the National Science Foundation of China (NSFC) under grant No. 61461136004, No. 61201191, No. 61321061, and No. 61401250, and Intel Collaborative Research Institute for Mobile Networking and Computing. }}

\maketitle

\begin{abstract}

To meet the surging demand of increasing mobile Internet traffic from diverse
applications while maintaining moderate energy cost, the radio access network
(RAN) of cellular systems needs to take a green path into the future,
and the key lies in providing elastic service to dynamic traffic
demands. To achieve this, it is time to rethink RAN architectures and expect
breakthroughs.  In this article, we review the state-of-art literature which
aims to renovate RANs from the perspectives of control-traffic decoupled air
interface, cloud-based RANs, and software-defined RANs. We then propose a
software-defined hyper-cellular architecture (SDHCA) that identifies a feasible
way of integrating the above three trends to enable green and elastic wireless
access. We further present key enabling technologies to realize
SDHCA, including separation of the air interface, green base station
operations, and base station functions virtualization, followed by our hardware testbed for SDHCA. Besides, we summarize
several future research issues worth investigating.

\end{abstract}

\section{Introduction}

Since their birth, cellular systems have evolved from the
first generation analog systems with very low data rate to today's fourth generation
(4G) systems with more than \SI{100}{Mbps} capacity to end users. However, the
radio access network (RAN) architecture has not experienced many changes:
base stations (BSs) are generally deployed and operated in a distributed fashion,
and their hardware and software are tightly coupled.
Facing the exponential growth of mobile Internet traffic on
one hand, and the significant energy consumption of mobile networks on the
other hand, breakthroughs are strongly expected in
RAN architecture design and corresponding systematic control methods. In the conventional RAN architecture,
even though there is little traffic requirement, the BSs cannot be switched off in order to maintain the basic coverage. This requires substantial static power to keep the BS active and to transmit the required signaling, thus causing energy waste.
To meet the urgent need for green and elastic wireless access, it is envisioned that
next-generation RANs should become increasingly software-defined
and the layout of their physical resources should break away from the fully distributed manner.

Along with the above paradigm shift, emerging RAN architectures have been developed
from three perspectives. The first is the new air interface architecture of
cellular networks that features signaling and data separation, aiming
at flexible and efficient control of small cells for throughput boosting and
BS sleeping-based energy
saving~\cite{niu2012energy,ishii2012novel}. To make the cell coverage more adaptive to the traffic
dynamics, some control signaling functions should be
decoupled from the data functions, and thus the data traffic service
is provided on demand while the control plane is always ``on'' to guarantee the
basic coverage. The second is renovating cellular networks into massive BSs with centralized
baseband processing and remote radio heads (RRHs), which further evolves to
have cloud-based baseband processing pool~\cite{cmri2013cran} and BS functions
virtualization. The third one is inspired by software-defined networking (SDN)
from wired networks, which separates the control and data
plane to enable centralized optimization of data forwarding.

We believe that successfully delivering green and elastic mobile access relies
on the deep convergence of the above three perspectives, of which the rationale is as follows. First, to realize the
control-traffic decoupled air interface,
flexible and efficient signal processing is required to reconstruct the frame
components from the control and traffic layers of the air interface.
The load of control signaling can also vary over time and space for nowadays mobile networks,
which requires the control coverage be reconfigurable with adaptive power and spectrum resources, in order to match the signaling load variations.
To tackle these challenges, cloud-based RAN architectures can offer help by providing
programmable BS functions and the reconfigurability of radio elements. Further,
the fronthaul network in cloud-based RAN architectures can be aided by SDN to
enable efficient data forwarding and flexible function splitting.
Meanwhile,
the control-data separation in SDN can be extended to the wireless access by
the control-traffic decoupled air interface.
Naturally, the air interface separation should converge with
cloud-based baseband processing under software-defined provisioning, exploiting its high flexibility and re-configurability. In fact, recent studies have begun to investigate
the integration of these perspectives~\cite{liu2014concert,zaidi2015future}.
However, the problems of how to combine the three
perspectives into a converged architecture and how to realize the architecture
in practice remain unclear.

To this end, in this article we propose a new software-defined
hyper-cellular architecture (SDHCA) which realizes the separation of the air
interface via a software-defined approach in a cloud-based infrastructure. We begin with reviewing the recent research on RAN architecture innovations from the
aforementioned three perspectives. Then we present the overall design of SDHCA that integrates air interface separation, cloud RAN and SDN, emphasizing the major technical contributions of SDHCA that bring elastic and green mobile service. In the next section, our initial research efforts towards the key enabling technologies of SDHCA are presented, followed by the testbed implementation showing the feasibility of SDHCA. Finally, we outlook future
research directions that can ultimately facilitate SDHCA into practical RANs.

\section{Recent RAN Architecture Developments}
\label{sec:review}

People have witnessed a rising interest in novel RAN architectures in recent literature.
We categorize them into three independent trends,
and summarize their main features and benefits in Table~\ref{tab:arches}.
The first trend is signaling-data separation at the air
interface. Among them the hyper-cellular architecture (HCA)~\cite{niu2012energy}
and the ``Phantom Cell'' concept~\cite{ishii2012novel} are typical examples.
Under such architectures, the network coverage is divided into two layers: control
coverage and traffic coverage. For instance in HCA, BSs are classified into two types: control base stations (CBSs)
and traffic base stations (TBSs).
Specifically, CBSs take care of control coverage, which provides network
access, system information broadcast and so on.
On the other hand, TBSs are meant for data traffic services to mobile users.
With the decoupled air interface, TBSs
can be switched on/off for significant energy savings without generating coverage holes.
Besides, CBSs can gather network control information and globally optimize the on/off states of TBSs and the radio resource allocation.
The idea of decoupled air interface has made its way into the cellular
standard, known as ``dual connectivity'' in 3GPP LTE Release 12~\cite{3gppdcr12}.

\begin{table}[!t]
  \centering
  \caption{Summary of new RAN architectures}
  \label{tab:arches}
  \begin{tabularx}{\textwidth}{|l|X|X|X|}
    \hline
    Trend & Literature & Features & Benefits \\
    \hline
    Decoupled air interface & \savespace
    \begin{compactitem}
    \item  HCN~\cite{niu2012energy}
    \item  Phantom Cell~\cite{ishii2012novel}
    \end{compactitem}
    & \savespace
    \begin{compactitem}
      \item Separation of control and data coverages
      \item CBSs gather network control information
    \end{compactitem} & \savespace
    \begin{compactitem}
      \item Energy saving
      \item Global optimization of network resources
    \end{compactitem} \\
    \hline
    Cloud-based RAN & \savespace
    \begin{compactitem}
    \item WNC~\cite{lin2010wireless}
    \item C-RAN~\cite{cmri2013cran}
    \end{compactitem} & \savespace
    \begin{compactitem}
      \item BBU RRH separation
      \item BBU consolidation
      \item Virtual base stations
    \end{compactitem} & \savespace
    \begin{compactitem}
      \item Cost reduction
      \item Improved flexibility
    \end{compactitem} \\
    \hline
    Software-defined RAN & \savespace
    \begin{compactitem}
    \item SoftRAN~\cite{gudipati2013softran}
    \end{compactitem} & \savespace
    \begin{compactitem}
      \item Control data separation
      \item Logically centralized controller
      \item Control APIs
    \end{compactitem} & \savespace
    \begin{compactitem}
      \item Global utility optimization
      \item Simplified network management
    \end{compactitem} \\
    \hline
    Integrated architectures &
    \begin{compactitem}
      \item OpenRAN~\cite{yang2013openran}
      \item CONCERT~\cite{liu2014concert}
      \item SDF~\cite{arslan2015software}
      \item Zaidi et al.~\cite{zaidi2015future}
    \end{compactitem} &
    \savespace
    \begin{compactitem}
      \item Integration of SDRAN and cloud computing
      \item Integration of SDRAN and decouple air interface
    \end{compactitem} & \savespace
    \begin{compactitem}
      \item Combined benefits
      \item Realization acceleration
    \end{compactitem} \\
    \hline
  \end{tabularx}
\end{table}

The industry and academia are also investigating integrating cloud computing
technologies into RANs. Among the most representatives are Wireless Network
Cloud~\cite{lin2010wireless} and Cloud RAN (C-RAN)~\cite{cmri2013cran}.
They share the same idea of consolidating base band units (BBUs) of BSs to a centralized computing cloud, while only leaving remote radio heads (RRHs) in the front end.
The cloud-based architecture can reduce the energy consumption, as well as the RAN
deployment and operational costs \cite{cmri2013cran}. Besides,
with virtualization technique, virtual base stations (VBSs) can be realized,
opening up the RAN for flexible system configurations and operations.

Thirdly, SDN has brought a rethinking of packet switching and routing in the Internet, and there have been emerging studies to
bring SDN concepts such as control-data separation, centralized control,
software applications programming interfaces (APIs) to RANs.
SoftRAN~\cite{gudipati2013softran} aims
to enable software-defined RANs (SDRANs). It introduces
the concept of big BS including a logically centralized controller
and distributed radio elements. With the centralized controller,
SDRAN optimizes the global utility over a local geographic area, and
simplifies the network management by software programming.

Recent literature has started to explore the integration of above trends for future RANs.
OpenRAN~\cite{yang2013openran} is proposed to utilize cloud computing resource pool and virtualization
to implement the SDRAN architecture.
CONCERT~\cite{liu2014concert} builds a converged cloud and cellular system based on control-data decoupling.
Arslan et al.~\cite{arslan2015software} propose the concept of
software-defined fronthaul (SDF) based on SDN and C-RAN.
Zaidi et al. propose an integrated architecture which combines SDN concepts
from SoftRAN and the signaling-data separation~\cite{zaidi2015future}.
These researches motivate us to explore the integration of the above three perspectives,
so as to make a further step to enable green and elastic wireless access in future
cellular systems.

\section{Software-Defined Hyper-Cellular Architecture}
\label{sec:SDHCA}

\begin{figure}[!t]
  \centering
  \includegraphics[width=0.93\textwidth]{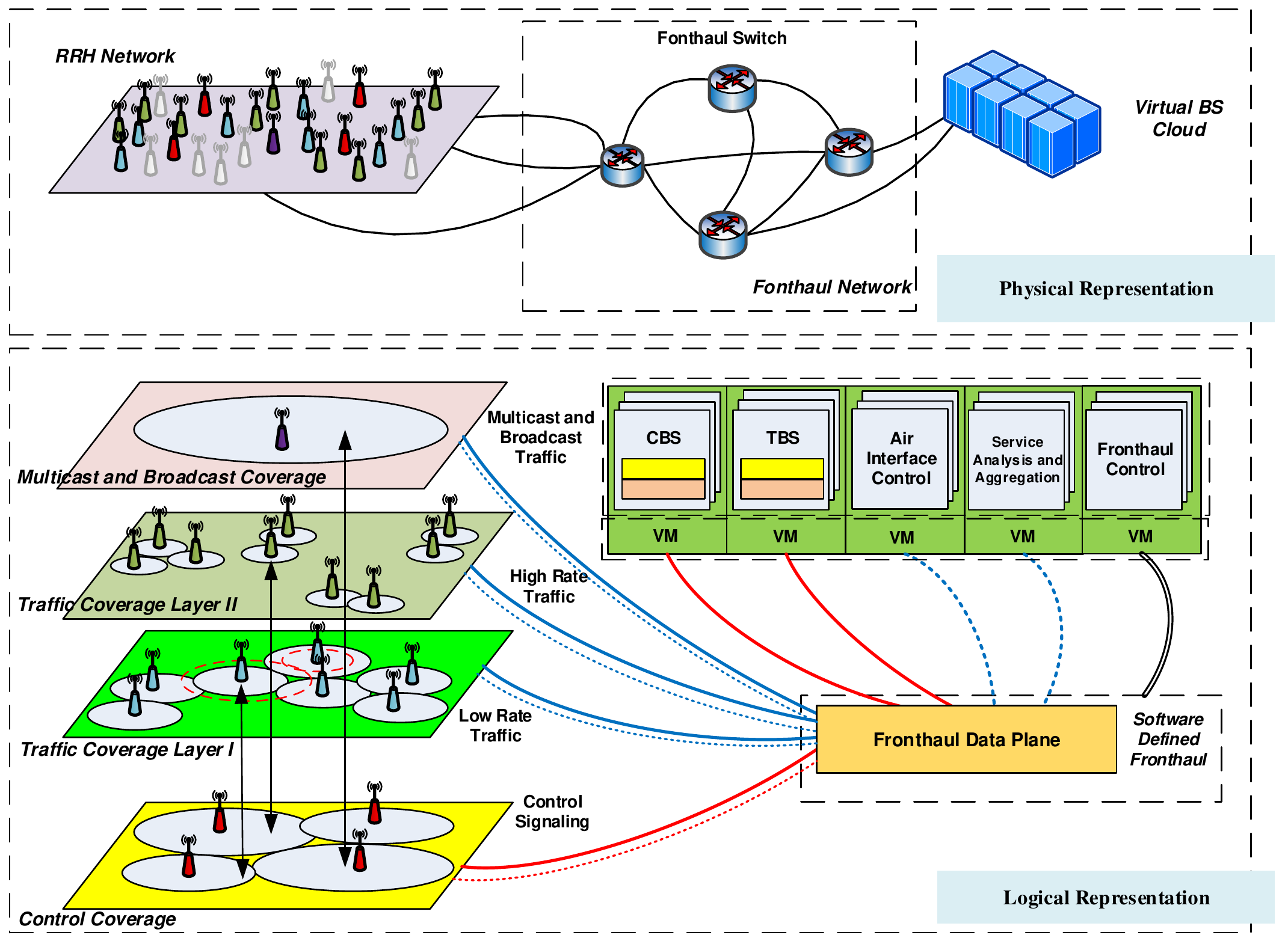}
  \caption{Software-defined hyper-cellular architecture.}
  \label{fig:arch}
\end{figure}

As shown in \figurename~\ref{fig:arch}, the SDHCA design is based on
the deep integration of air interface separation, cloud RAN and SDN. It exploits the cloud infrastructure, which can be
divided into three subsystems: the RRH network, the fronthaul network, and the
virtual BS (VBS) cloud. The radio elements (RRHs) can be as simple as merely dealing with RF transmission/reception, or having some baseband processing functions, and their roles can be dynamically
configured as CBS or TBS, or put into sleep mode, according to the network status and hardware capabilities of the RRHs. The deployment of the RRHs can be done via conventional network planning mechanisms, satisfying the peak-hour traffic of the network.
From the perspective of logical functions,
the proposed SDHCA provides one control coverage layer and
multiple conceptual layers for different user traffic types. In this way, the cell
coverage is ``softer'' and smarter to deliver greener wireless access.
RRHs are connected to the VBS cloud via the fronthaul network, which
is also software defined. In the cloud, the functions of CBS and TBS are realized as VBS applications in
virtual machines (VMs).

The key features of SDHCA are summarized as follows:
\begin{itemize}
  \item \emph{Control data separation.} The separation lies in three aspects. First,
    in the air interface, CBSs are in charge of control coverage, while TBSs
    are responsible for traffic coverage. Second, on the infrastructure level,
    the software in charge of the network functions is separated from the
    hardware which forwards or transmits the data. In particular, one RRH can
    be dynamically configured to act as a CBS or a TBS or even both, i.e., handling some traffic while acting as a CBS. Last but not
    least, the control plane and the data plane of the SDF network are also
    decoupled.
  \item \emph{CBS as the RAN controller.} CBSs take care of mobile users and
    TBSs underlying their coverage. Besides, CBSs also control the fronthaul network.
    In this way, the CBS holds a
    global view of the RAN in a local geographic area, and optimizes the
    on-demand configuration and activation of TBSs, so that the network resources,
    including spectrum resources and energy resources, can well match the
    dynamic traffic in an elastic way. When the traffic load changes, CBS can also control the cell zooming behaviour of active TBSs to balance the
load, as shown in the red dashed circles on the traffic layer I in \figurename~\ref{fig:arch}.
  \item \emph{Software-defined network functions via virtualization.} The network
    functions, including air interface control, service analysis and
    aggregation, baseband sample generation, the fronthaul control plane and so on, are realized by software
    applications running in VMs. The functions are thus easily
    programmed and updated, allowing for flexible and efficient network
    operations, potentially reducing the computing energy consumption.
\end{itemize}

Thanks to the above tightly integrated features, SDHCA provides flexible services to users exploiting spatial-temporal variations of the traffic demand, so that the energy efficiency of the whole network can be greatly improved:
\begin{itemize}
\item The decoupled air interface produces flexible sleeping opportunities for TBSs. As shown in \figurename~\ref{fig:arch}, the access control and other coverage-related
signaling of different traffic layers are handled by CBSs, and the vertical
arrows in \figurename~\ref{fig:arch} indicate that cells in the traffic layers are covered by
a CBS in the control layer based on their positions. Therefore, the RRHs can
be configured to TBSs on any traffic layer shown in \figurename~\ref{fig:arch}, and they can be switched off without having coverage holes, leading to significant energy saving. Regarding active TBSs, the energy consumption related to control signaling can also be saved.
\item The required compute resources in the BBU pool can adapt to the number of active RRHs, and so do the virtualized fronthauling functions needed to support these active RRHs. For instance, the VMs that run the CBS and TBS realizations can be constructed or released on demand based on the active/sleeping status of the associated CBSs and TBSs, saving the energy of the computing cloud.
\item Unlike conventional cost-inefficient solutions that rely on dark fibers to connect RRHs to the VBS cloud, the SDF enables flexible mapping between BBUs and RRHs~\cite{arslan2015software}, and efficient baseband function splitting~\cite{liu2015graph} for equivalent baseband signal compression, so that the high bandwidth requirement of the fronthaul is guaranteed with low cost.
\end{itemize}
In short, the proposed SDHCA is a viable solution to offer green and elastic mobile access.

\section{Enabling Technologies}
\label{sec:key}

The enabling technologies of SDHCA are described in details in this section. We first compare existing air interface separation solutions and discuss the future separation-oriented air interface design. Then our initial research efforts to realize green BS operations with BS dispatching and BS sleeping are presented. Finally, possible solutions of BS functions virtualization in SDHCA are discussed.

\subsection{Separation of Air Interface}

\begin{table}[!t]
  \centering
  \caption{Comparison of air interface separation schemes}
  \label{tab:sep}
  \begin{tabularx}{\textwidth}{|X|X|X|}
    \hline
    Scheme & Advantages & Drawbacks \\
    \hline
    \savespace
    Extreme separation:
    \begin{compactitem}
    \item TBS: User data
    \item CBS: All others (including pilot)
    \end{compactitem}
    & -- & Unsuitable \\
    \hline
    \savespace
    Functionality separation~\cite{xu2013functionality}:
    \begin{compactitem}
    \item CBS: Synchronization, broadcast of system information, paging, multicast
    \item TBS: Synchronization, unicast
    \end{compactitem}
    & Analytic feasibility
    validation & LTE specific \\
    \hline
    \savespace
    HyCell~\cite{zhao2015hycell}
    \begin{compactitem}
    \item Joint functionality and logical channel separation
    \item Synchronization only at CBS
    \end{compactitem}
    & \savespace
    \begin{compactitem}
    \item Generic to multiple standards
    \item Testbed evaluation
    \end{compactitem}
    & Evaluation only on GSM/GPRS \\
    \hline
    \savespace
    Dual connectivity~\cite{3gppdcr12}:
    \begin{compactitem}
    \item C-Plane: CBS sends out RRC
    \item U-Plane:
      \begin{compactitem}
      \item Option 1: Independent PDCP
      \item Option 2: Slave bearer PDCP at CBS, RLC  and lower at TBS
      \end{compactitem}
    \end{compactitem}
    & \savespace
    \begin{compactitem}
    \item Standard
    \item Minimal change
    \end{compactitem}
    & Suboptimal \\
    \hline
    \savespace
    Separation-oriented air interface:
    \begin{compactitem}
    \item Separation in design phase
    \item Independent optimization
    \item Software defined
    \end{compactitem}
    & \savespace
    \begin{compactitem}
    \item Simple and efficient
    \item Programmable
    \end{compactitem} &
    Backward compatibility breakage \\
    \hline
  \end{tabularx}
\end{table}

For possible air interface separation schemes, we observe that
the naive ``extreme separation'' scheme is in fact unsuitable.
Here ``extreme separation'' means only the transmission of user perceived data is
handled by the TBS while all other parts are processed at the CBS.
The reason why extreme separation fails is that modern cellular systems rely on
pilot symbols to estimate the wireless channel to aid the decoding of user data
bits. With pilots transmitted by the CBS and data bits transmitted from the TBS, this extreme separation approach will lead to inaccurate
channel estimation and thus incorrect decoding of the data.

One possible way suggested in the literature is functionality
separation~\cite{xu2013functionality}, where functionality is defined as the essential
sets of functions provided by the network to the mobile users.
It can be divided into five classes: synchronization, broadcast of system
information, paging, multicast (for low-rate data transmissions such as voice), and
unicast (for high-rate data transmissions).
Through the functionality separation,
the CBS is responsible for the former four classes, and the TBS for synchronization and high-rate data transmission.
Careful analysis was conducted to ensure the user equipment (UE) state
transitions  continue to work after the separation, although specific to the LTE standard.
Based on functionality separation, an alternative separation scheme was proposed
in HyCell~\cite{zhao2015hycell}, which will be discussed in details in the next section.

Besides, the separation of the air interface has also undergone discussions in 3GPP,
termed as ``dual connectivity''~\cite{3gppdcr12}.
In dual connectivity, CBS and TBS are called as PeNB and SeNB respectively, and
the separation scheme is specified in the aspects of control plane (C-Plane)
and user plane (U-Plane).
In C-Plane, only CBS sends out radio resource control (RRC) messages to the user after coordination with the TBS.
Two options are available for the U-Plane:
1) Independent packet data convergence protocol (PDCP) layers are used at the CBS and the TBS, and each node carries one
bearer;
2) Slave bearer is split between PDCP and radio link control (RLC) layers, so that CBS processes the PDCP
layer and TBS processes RLC and lower layers.
The advantage of dual connectivity is that it requires minimum
changes to the overall RAN architecture in LTE, and one can expect
standard terminals to gradually support it.
However, minimum changes also limit the degree of freedom in separation, which
implies the separation scheme can be suboptimal.

For future cellular systems, a new separation-oriented air interface design is more preferable
if one is allowed to break backward compatibility with earlier cellular standards, guided by the principle of separating control from traffic.
It is expected to reduce redundancy and keep the protocol simple. It can also improve the network efficiency through independent optimization of
the air interface at the CBS and the TBS according to their different characteristics.
Besides, the air interface should be made easily programmable and upgradable
via software, thus improving the flexibility. The discussion of the aforementioned separation schemes are summarized in Table~\ref{tab:sep}.

\subsection{Green Base Station Operations}

\begin{figure}[!t]
\centering
\subfloat[]{\includegraphics[width=3.1in]{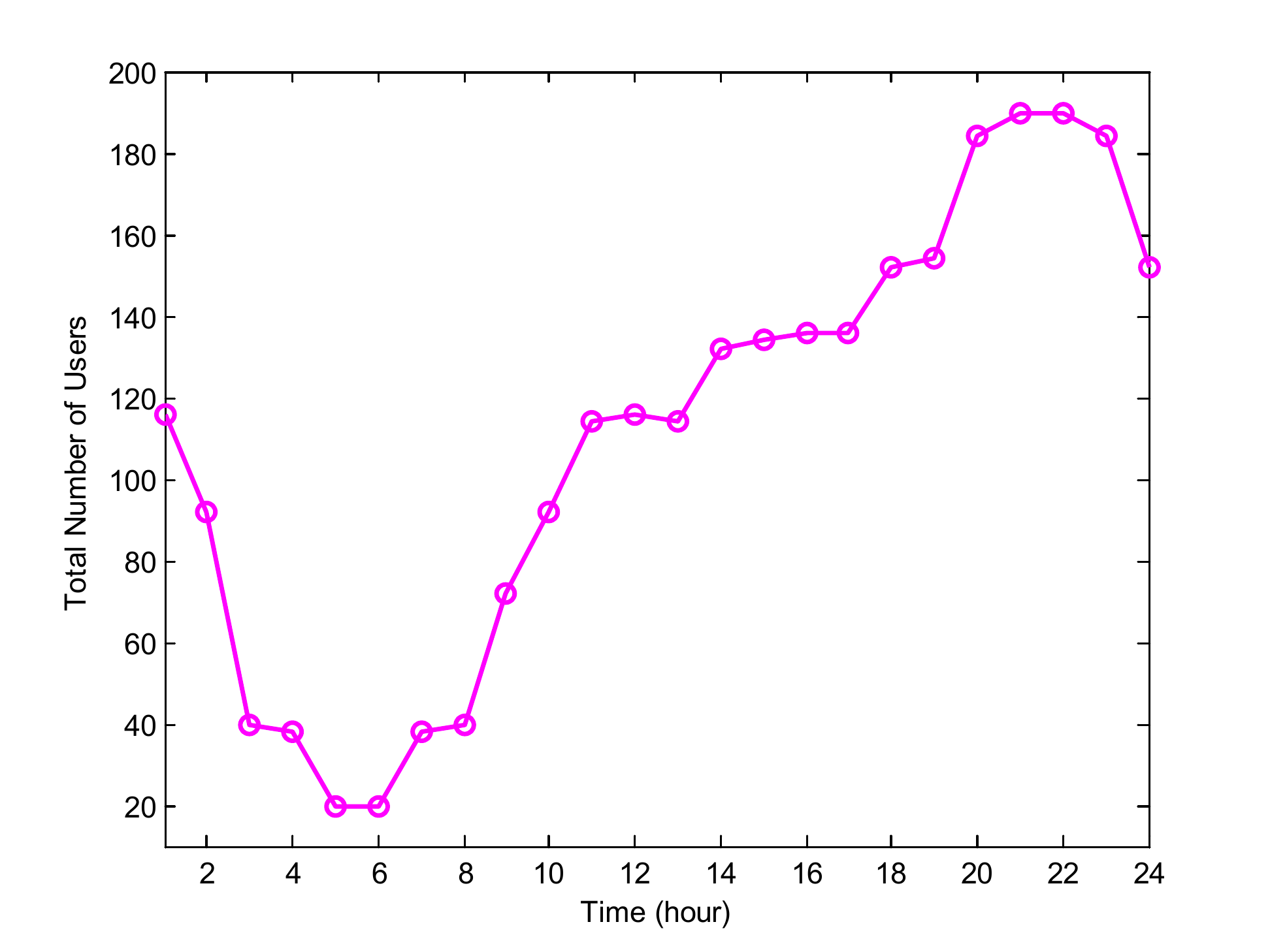}
\label{fig_first_case}}
\hfil
\subfloat[]{\includegraphics[width=3.1in]{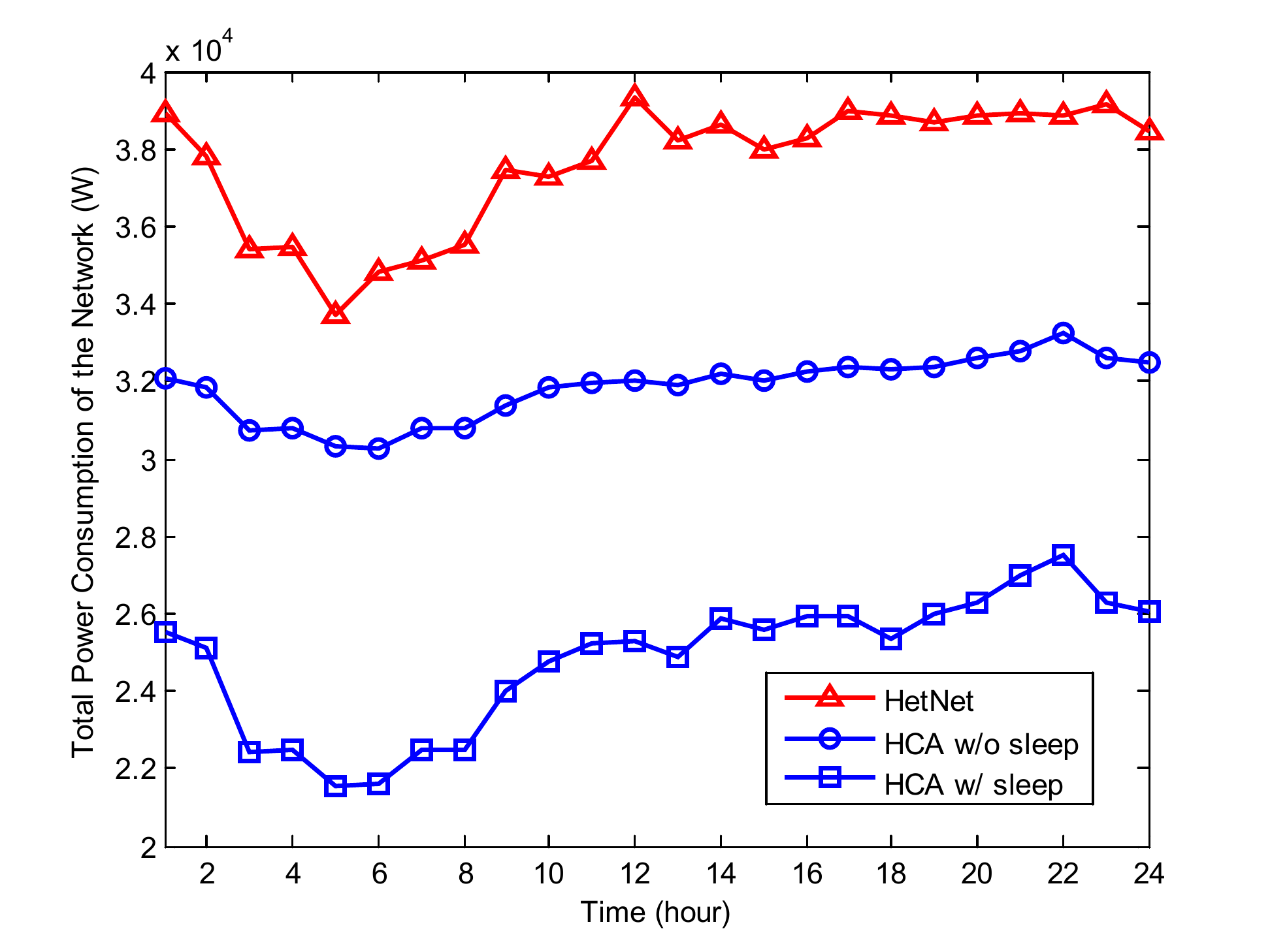}
\label{fig_second_case}}
\caption{Simulation comparison of the network power consumption. (a) Average total number of users in the network for each hour; (b) Average total network power consumption for each hour.}
\label{fig_sim}
\end{figure}

As mentioned in Section \ref{sec:SDHCA}, SDHCA can improve the network energy efficiency by allowing flexible TBS sleeping and reducing the signaling related energy consumption on TBSs. We
conduct a simulation study to evaluate the energy saving gain.  We use a system level simulator in accordance with
3GPP simulation requirements, and the BS power model is from the EARTH project
\cite{Earth}. The layout of 19 CBSs follows a standard hexagonal topology, while
the distribution of TBSs follows a homogeneous Poisson point process with
totally 38 TBSs on average. Users, each with an FTP downloading source of the same volume, are randomly and uniformly dropped in the 19-hexagonal area, and we vary the number of users according to a daily traffic
profile provided by EARTH \cite{Earth}, resulting in the average total number of users in
each hour as shown in \figurename~\ref{fig_first_case}. In \figurename~\ref{fig_second_case}, three schemes are compared. The ``HCA w/o sleep" corresponds to the case when TBSs are not allowed to sleep, while in ``HCA w/ sleep" TBSs can dynamically go to sleep according to a simple load-threshold-based policy. The ``HetNet" corresponds to conventional heterogeneous network, where we turn CBSs into macro BSs and TBSs into micro BSs, from ``HCA w/o sleep". The time-frequency resource blocks (RBs) are fully used in all BSs in ``HetNet", while on CBSs, only those for control signaling are used, and the RBs for control signaling are muted on TBSs. The energy reduction from ``HetNet" to ``HCA w/o sleep" is due to the muted control-signaling-related RBs, while the energy saving from ``HCA w/o sleep" to ``HCA w/ sleep" is due to the dynamic TBS sleeping. Note that the separated air interface guarantees this flexible TBS sleeping.

One of the fundamental issues in SDHCA or any air interface separation scheme is the channel
condition acquisition, which is particularly important for BS dispatching and
sleeping. When a new user arrives, the
network should dispatch the best RRH to act as its serving TBS, possibly
requires waking up a sleeping RRH. However, getting to know the channel
conditions of the sleeping RRH to the user is very challenging.

\begin{table}[!t]
\centering
\caption{Prediction accuracy / relative prediction error of different
algorithms.}
\label{tab:pred}
\begin{tabular}{|c|c|c|c|c|c|c|c|}
\hline
No. of Scatterers & 10 & 15 & 20 & 25 & 30 & 35 \\
\hline
RS & 0.2/0.4062 & 0.2/0.3815 & 0.2/0.3665 & 0.2/0.3562 & 0.2/0.3488 & 0.2/0.3395 \\
\hline
KNN ($K=5$) & 0.5601/0.1741 &	0.5532/0.1622 &	0.5420/0.1606 &	0.5306/0.1601 &	0.5188/0.1593 &	0.5029/0.1563 \\
\hline
NN-CR & 0.7483/0.0662 &	0.7249/0.0703 &	0.7140/0.0662 &	0.6871/0.0744 &	0.6700/0.0791 &	0.6528/0.0855 \\
\hline
NN-LO & 0.8268/0.0317 &	0.7972/0.0376 &	0.7845/0.0392 &	0.7788/0.0398 &	0.7584/0.0429 &	0.7443/0.0476 \\
\hline
\end{tabular}
\end{table}

We thus propose a novel way based on machine learning to see the unobservable
channel conditions of the sleeping RRHs. The main idea is to build a mapping
function from the user's observable channel conditions of active CBSs and TBSs
to the unobservable channel conditions of sleeping RRHs.
A neural network (NN)-based algorithm for RRH selection
learning in SDHCA is designed, which combines the standard
approaches in NN with crafted processing procedures including discrete Fourier
transform (DFT), quantization by
logarithmic treatment and the Lloyds algorithm to form input features. We consider a single CBS with 80
antennas and 5 candidate single-antenna RRHs in sleep. The objective is to
select a RRH from the 5 candidates with the best channel gain to a user, given that the
user only has the instantaneous channel condition to the CBS, and historical
channel conditions to the CBS and RRHs recorded by other users. The prediction
accuracy is defined as the percentage of the correct selection. Another
performance metric is relative selection error between the predicted RRH's
channel gain and the actual best channel gain. We compare our algorithm with
other algorithms and the results are shown in Table~\ref{tab:pred}. The random selection (RS)
method provides the baseline accuracy, which is 20\% for all scatterer
configurations. The simple K-nearest-neighbor (KNN) algorithm, which outputs
the dominant choice among the $K$ nearest neighbors in the channel space,
increases the accuracy to about 55\%. In comparison, the accuracy of the
proposed NN-based channel learning algorithm with channel response as input (NN-CR) is around 70\%. Moreover, there is an 8\% gap
between the accuracies of NN-CR and a NN-based algorithm using
genuine user location as inputs (NN-LO). But note that in practical systems,
location information sufficiently accurate for channel estimation is generally hard to
obtain.

\subsection{Base Station Functions Virtualization}

BS functions virtualization is part of Network Functions
Virtualization (NFV) in cellular systems. Proposed and standardized by ETSI,
NFV aims to virtualize the network node
functions and build virtual networks. With BS functions virtualization, BSs
become VBSs, and their functions are software defined and can provide various APIs to
network operators.

Currently there are several proof-of-concept implementations of
VBSs in cellular networks, typically implemented on hypervisor-based virtualization platforms,
such as KVM and VMWare ESX, and one major challenge is providing realtime performance~\cite{cmri2013cran}.
Modern cellular systems have stringent realtime
requirements. For example, the LTE standard specifies that one subframe must be acknowledged
after three subframes upon reception in frequency division duplex (FDD) mode.
It leaves a total of \SI{3}{ms} budget for decoding and subframe generation.
To fulfill the task, software optimization and hardware accelerators are
employed in current implementations. In the software domain, realtime optimization of the whole software stack,
including the host and guest operating system (OS, typically Linux) kernel, the hypervisor,
and the guest applications, are necessary to power up modern cellular
standards on general purpose processor (GPP) platforms, which makes BS
functions virtualization a daunting task.

Container virtualization has the potential to provide a lightweight yet effective way to
virtualize the BS functions and realize SDHCA.
Compared with hypervisor-based virtualization, container virtualization
eliminates the need of a guest OS. By reducing the intermediate virtualization
layers, container virtualization provides better performance than hypervisor-based
virtualization.
Therefore, when using container virtualization to build VBSs, the need of realtime optimization can be relaxed.

Another issue of realizing BS functions on virtualization is
inter-VM communication.  Virtualization typically provides isolation across VMs
for fault tolerance, but it also makes inter-application interactions difficult, because the network
communication across VMs can degrade the system performance
dramatically.  However, CoMP (Coordinated Multi-Point Communication) and
CBS-TBS signaling rely on efficient inter-VBS communications.  To
tackle this issue, new mechanisms are needed to share resources such as disk, memory, or CPU cache, to facilitate inter-VBS communications.
Instruction set architecture enhancements can also be exploited as a
complementary approach to improve the performance of VM communications.

\section{Hardware Testbed}
\label{sec:testbed}

\begin{figure}[!t]
  \centering
  \includegraphics[width=\textwidth]{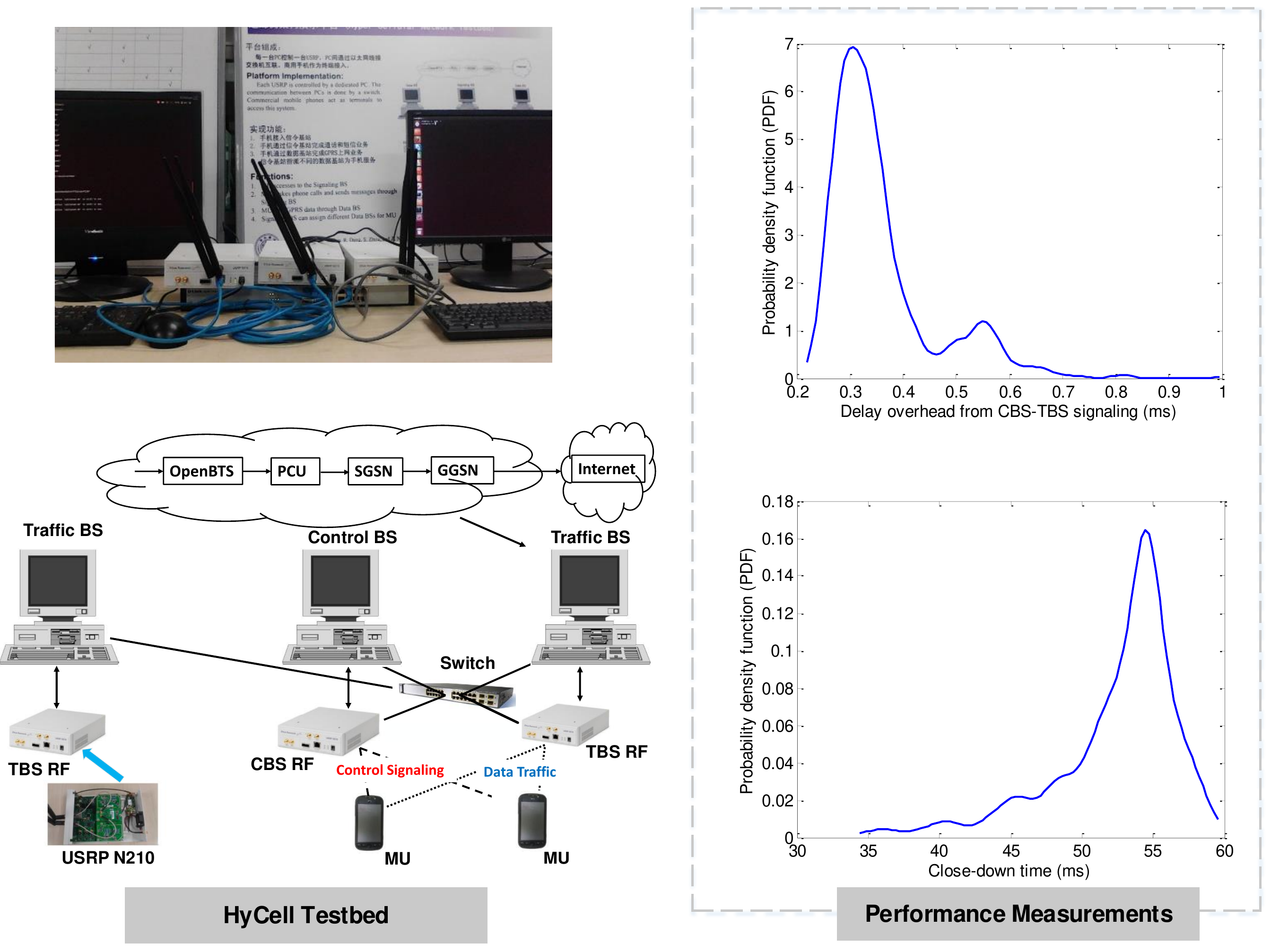}
  \caption{The HyCell testbed.}
  \label{fig:testbed}
\end{figure}

We are prototyping hardware testbed for systematic evaluation of SDHCA, which
is realized via programmable software on GPP platforms.
Fig.~\ref{fig:testbed} shows our current testbed called
HyCell~\cite{zhao2015hycell}, which is
based on OpenBTS%
\footnote{\url{http://openbts.org/}}, an open source GSM/GPRS BS application,
and the USRP \footnote{\url{http://www.ettus.com//}} hardware platform.
The overall system structure including hardware interconnection and major
software modules are shown in the left part of Fig.~\ref{fig:testbed}. In the testbed, the association between the BS servers and the USRP devices is static due to hardware limitations. The static assignment makes it relatively simple and easy to implement our testbed, while it cannot fully reveal SDHCA's potential of adaptiveness to traffic dynamics.

Our testbed work presents an alternative separation scheme by jointly
considering the functionality and logical channels of current standards. In
this design, synchronization only resides at the CBS side, and the TBS is only
in charge of high-rate data transmission. As a result, synchronization between
the TBS and the UE is guaranteed by CBS-UE synchronization over the air.
One major advantage of this synchronization scheme is that no user side modification is needed,
so that existing mobile terminals can seamlessly access the renovated cellular network.
The scheme has been evaluated on our GSM/GPRS based testbed implementation,
and is generic to multiple standards.
Currently we are investigating its implementation over the LTE standard based on
an open source platform OpenAirInterface%
\footnote{\url{http://www.openairinterface.org/}}.

Besides, our testbed is able to demonstrate green BS operations including BS
dispatching and BS sleeping. BS sleeping is utilized when the network load is light, and
some TBSs can be switched off for energy saving, while load-balancing BS dispatching is beneficial for highly-loaded network or when some TBSs are in sleep mode and the remaining ``active" network is of high load.
In the testbed, we also evaluate the delay overhead of the air interface separation and BS sleeping scheme, which are important factors to guarantee the quality of service (QoS) while enjoying the energy saving benefits.
In the performance measurements of Fig.~\ref{fig:testbed}, we have shown the delay overhead of signaling interaction between the CBS and the TBS when processing the channel request from a user terminal, of which the mean is about \SI{0.36}{ms} and the standard variance is about \SI{0.1}{ms}. The small values indicate the feasibility of the air interface separation in SDHCA. As for BS sleeping, we measure the close-down time of turning off a TBS, and its mean is about \SI{52}{ms}. On the other hand, the set-up time of waking up a TBS is of several seconds, whose optimization is needed in future study.

Regarding BS functions virtualization, we are prototyping a VBS pool on a commodity x86 server based on
Docker\footnote{\url{https://www.docker.com/}}
containers, so as to implement dynamical resource allocation algorithms for multiple
VBSs and design the API for software-defined network control and management, including more sophisticated BS dispatching and sleeping algorithms.
We envision a complete SDHCA testbed to demonstrate its
concepts, compare different algorithms, and evaluate system performance
as well as inspire new research directions.

\section{The Way Forward}

Synchronization is a big challenge to realize SDHCA. Fine synchronization
between different network elements is required to enable green BS operations.
Current implementations typically make heavy use of GPS receivers. However,
it incurs additional cost, and might not work well in indoor and underground
environments. Alternative solutions such as fronthaul-based synchronization or
radio-based synchronization should be investigated to tackle this challenge.

Inherited from the cloud-based RAN, SDHCA also faces the challenge of designing high-bandwidth fronthaul. Moreover, future fronthaul networks can have heterogeneous physical realizations, including wireless, fiber, high-speed Ethernet, and so on. As a promising solution of the next generation fronthaul, software-defined packet switching should be evaluated and how to realize it needs to be answered \cite{Liu2015FHN}.

From the perspective of better satisfying mobile users' needs,
SDHCA should be programmed to provide user-centric services
by making the network aware of mobile users' states and allocating network
resources to deliver their request contents.
How to capture users' state and how to schedule users accordingly are yet to be
addressed, and for example, interference-aware RRH selection in SDHCA is a valuable yet challenging issue.

By building virtual network functions on a software-defined platform, the same
physical infrastructure can be shared by multiple network operators, including
virtual operators. Dynamic RAN sharing relies on high level abstraction of RAN
as a controllable entity. Technical problems lie ahead including RAN resource
slicing and isolation, as well as RAN provisioning and orchestration. Besides,
novel business models between multiple operators call for investigation,
possibly via a game theoretic approach.

SDHCA enables green and elastic
wireless access, and can further serve as a platform for RAN innovations.
In this work we focuses on the RAN part of cellular systems. There are quite a
few works trying to bring SDN and NFV concepts to the core network part.
How to combine the RAN innovations with core network advances is an exciting
research direction.


\begin{thebibliography}{99}
\providecommand{\url}[1]{#1}
\csname url@samestyle\endcsname
\providecommand{\newblock}{\relax}
\providecommand{\bibinfo}[2]{#2}
\providecommand{\BIBentrySTDinterwordspacing}{\spaceskip=0pt\relax}
\providecommand{\BIBentryALTinterwordstretchfactor}{4}
\providecommand{\BIBentryALTinterwordspacing}{\spaceskip=\fontdimen2\font plus
\BIBentryALTinterwordstretchfactor\fontdimen3\font minus
  \fontdimen4\font\relax}
\providecommand{\BIBforeignlanguage}[2]{{%
\expandafter\ifx\csname l@#1\endcsname\relax
\typeout{** WARNING: IEEEtran.bst: No hyphenation pattern has been}%
\typeout{** loaded for the language `#1'. Using the pattern for}%
\typeout{** the default language instead.}%
\else
\language=\csname l@#1\endcsname
\fi
#2}}
\providecommand{\BIBdecl}{\relax}
\BIBdecl

\bibitem{niu2012energy}
Z.~Niu, S.~Zhou, S.~Zhou, X.~Zhong, and J.~Wang, ``Energy efficiency and
  resource optimized hyper-cellular mobile communication system architecture
  and its technical challenges,'' \emph{SCIENTIA SINICA Informationis},
  vol.~42, no.~10, pp. 1191--1203, 2012, (In Chinese).

\bibitem{ishii2012novel}
H.~Ishii, Y.~Kishiyama, and H.~Takahashi, ``A novel architecture for {LTE-B}:
  {C}-plane/{U}-plane split and {P}hantom {C}ell concept,'' in \emph{IEEE
  GLOBECOM Int. Workshop Emerging Technologies for LTE-Advanced and Beyond-4G},
  Dec. 2012, pp. 624--630.

\bibitem{cmri2013cran}
Ch.-L.~I, J.~Huang, R.~Duan, C.~Cui, J.X.~Jiang, L.~Li, ``Recent progress on C-RAN centralization and cloudification," \emph{IEEE Access}, vol.~2, pp.~1030--1039, 2014.

\bibitem{yang2013openran}
M.~Yang, Y.~Li, D.~Jin, L.~Su, S.~Ma, and L.~Zeng, ``{O}pen{RAN}: A
  software-defined {RAN} architecture via virtualization,'' in \emph{Proc. ACM
  SIGCOMM 2013 Conf.}, Hong Kong, China, Aug. 2013, pp. 549--550.

\bibitem{liu2014concert}
J.~Liu, T.~Zhao, S.~Zhou, Y.~Cheng, and Z.~Niu, ``{CONCERT}: A cloud-based
  architecture for next-generation cellular systems,'' \emph{{IEEE} Wireless
  Commun. Mag.}, vol.~21, no.~6, pp. 14--22, Dec. 2014.

\bibitem{3gppdcr12}
\BIBentryALTinterwordspacing
3GPP, ``{3GPP} {TR} 36.842: Study on small cell enhancements for {E-UTRA} and
  {E-UTRAN} --higher layer aspects,'' Dec. 2013, (version 12.0.0 Release 12).
  [Online]. Available:
  \url{http://www.3gpp.org/ftp/Specs/archive/36_series/36.842/36842-c00.zip}
\BIBentrySTDinterwordspacing

\bibitem{lin2010wireless}
Y.~Lin, L.~Shao, Z.~Zhu, Q.~Wang, and R.~Sabhikhi, ``Wireless network cloud:
  Architecture and system requirements,'' \emph{IBM Journal of Research and
  Development}, vol.~54, no.~1, pp. 4:1--4:12, 2010.

\bibitem{gudipati2013softran}
A.~Gudipati, D.~Perry, L.~E. Li, and S.~Katti, ``{S}oft{RAN}: Software defined
  radio access network,'' in \emph{Proc. 2nd ACM SIGCOMM Workshop Hot Topics in
  Software Defined Networking (HotSDN '13)}, Hong Kong, China, Aug. 2013, pp.
  25--30.

\bibitem{arslan2015software}
M.~Y. Arslan, K.~Sundaresan, and S.~Rangarajan, ``Software-defined networking
  in cellular radio access networks: Potential and challenges,'' \emph{{IEEE}
  Commun. Mag.}, vol.~53, no.~1, pp. 150--156, Jan. 2015.

\bibitem{zaidi2015future}
Z.~Zaidi, V.~Friderikos, and M.~Imran, ``Future {RAN} architecture: {SD-RAN}
  through a general-purpose processing platform,'' \emph{{IEEE} Veh. Technol.
  Mag.}, vol.~10, no.~1, pp. 52--60, Mar. 2015.

\bibitem{liu2015graph}
J.~Liu, S.~Zhou, J.~Gong, Z.~Niu, and S.~Xu, ``Graph-based framework for
  flexible baseband function splitting and placement in C-RAN,'' in \emph{IEEE
  ICC 2015}, London, United Kingdom, Jun. 2015, pp. 958--1963.

\bibitem{xu2013functionality}
X.~Xu, G.~He, S.~Zhang, Y.~Chen, and S.~Xu, ``On functionality separation for
  green mobile networks: Concept study over {LTE},'' \emph{{IEEE} Commun.
  Mag.}, vol.~51, no.~5, pp. 82--90, 2013.

\bibitem{Earth}
G.~Auer, et al. ``How much energy is needed to run a wireless
  network?" \emph{IEEE Wireless Commun.}, vol.~18, no.~5, pp.40--49, 2013.

\bibitem{zhao2015hycell}
T.~Zhao, L.~Wang, X.~Zheng, S.~Zhou, and Z.~Niu, ``{H}y{C}ell: Enabling {GREEN}
  base station operations in software-defined radio access networks,'' in
  \emph{IEEE ICC 2015 Workshop Next Generation Green ICT}, London, United
  Kingdom, Jun. 2015, pp. 2868--2873.

\bibitem{Liu2015FHN}
J.~Liu, S.~Xu, S.~Zhou, and Z.~Niu, ``Redesigning fronthaul for next-generation networks: beyond baseband samples and point-to-point links, " \emph{IEEE Wireless Commun.}, vol.~22, no.~5, Oct. 2015.

\end{thebibliography}
\end{document}